\documentstyle[12pt]{article}
\normalbaselines
\catcode `\@=11
\@addtoreset{equation}{section}

\def\section{\@startsection {section}{1}{\z@}{-3.5ex plus -1ex minus
     -.2ex}{2.3ex plus .2ex}{\normalsize\bf}}
\def\subsection{\@startsection{subsection}{2}{\z@}{-3.25ex plus -1ex minus
-.2ex}{1.5ex plus .2ex}{\normalsize\bf}}

\def\thebibliography#1{\section*{References}
\list
  {[\arabic{enumi}]}{\settowidth\labelwidth{[#1]}\leftmargin\labelwidth
  \advance\leftmargin\labelsep
  \usecounter{enumi}}
  \def\newblock{\hskip .11em plus .33em minus -.07em}
  \sloppy
  \sfcode`\.=1000\relax}

\catcode `\@=12

\begin{document}

\begin{titlepage}
\vspace*{2.5cm}

\begin{center}

{\bf THE NEUTRINOS, THE GRAVITATION AND
     THE GAUGE FORMALISM}\vspace{1.3cm}\\
\medskip

{\bf Valery Koryukin}
\vspace{0.3cm}\\

 Department of Applied Mathematics, Mari State Technical University\\
 Box 179, Main Post--Office, Yoshkar--Ola, 424000, Russia\\
 e-mail: koryukin@mpicnit.mari.su

\end{center}

\vspace*{0.5cm}

 The detection and the research of the neutrinos background of
 Universe are the attractive problems. This problems do not seem
 the unpromising one in the case of the high neutrinos density
 of Universe. It was offered before to use the low energy
 neutrinos background of Universe for the explanation of the
 gravitational phenomena with the quantum position attracting
 the Casimir's effect for this. As a result it was connected
 the gravitational constant with the parameters characterizing
 the electroweak interactions. If now we shall be based on
 the results of the experements fixing the equality of
 the gravitation mass and the inert one then it can
 consider that the spectrum of the particle masses is defined by
 their interaction with the neutrinos background of Universe. This
 statement is confirmed what the rest mass of the photon is equal to
 zero in contradistinction to the masses of the vector
 bosons~$W^{+}$, $W^{-}$, $Z^{0}$ whiches interact with the
 neutrinos immediately.



\hspace{2mm}
\end{titlepage}

\section{\hspace{-4mm}.\hspace{2mm} The Casimir's effect
 and the gravitation}

The making of a physical theory embracing all an energy
spectrum of interactions is a fairly difficult task.
In consequence a construction of asymptotical theories both in
high-energy and low-energy ends of this spectrum was
justified historically. The most considerable success
attended the work in the high-energy approximation, in a
result of which was made quantum electrodynamics (QED)
giving the prediction confirming
experimentally with the remarkable precision. Naturally, that
this theory became the imitation specimen by the construction
of the analogous theories of the strong interaction (the
quantum chromodynamics (QCD)) and the weak one (Salam-Weinberg
model). The every possible theories describing continuum with
the large number of particles such as the theories of a solid
body, a liquid, a gas, a plasma, an electromagnetic radiation,
shells, nuclei at low energies and also General Relativity (GR)
is related to the opposite end of the energy spectrum. GR look
the exclusion against this background, what gave the reason to
consider the gravitation is only the effect of the existence of
the space-time curvature.

Admittedly the  theories wrecking the present idea are
appearing in the second half of our century such as the two-tensor
gravitation theory~\cite{sa}, in which was made the attempt to
rewrite the theory of nuclear interactions into the geometrical
language. It can attribute to like works also and the gauge
theory of the dislocations and disclinations~\cite{ce}. In consequence
of this the transfer to the geometrization description as the
most comfortable one in the long-wave length range for any
interactions is the logical one. Thereby it is wrecked fully the
exceptionality of the gravitation and the forces corresponding
to it are not to be distinguished between others, such as
Yukawa's forces and the Van der Waals' forces. So it is necessary
to show that the gravitation interaction is not a fundamental
one, but the one is induced by others interactions as possible
hypothetical ones. The more so, that the gravitational
constant~$G_N \approx 6.7 \cdot 10^{-39}~GeV^{-2}$ (it is used
the system of units~$\hbar =c=1$,
where~$2 \pi \hbar$ is the Planck's constant and~$c$ is the light
speed) is a suspicionsly small value and a dimensional one
furthermore (as is known the latter prevent to the constraction of
the renormalizable quantum theory).

Before building the theory of the induced gravitation on the base
of the hypothetical interactions and the hypothetical particles
it was necessary to verify the possibility of the utilization of
the known particles and the known interactions for this purpose.
Naturally that the neutrinos are the most suitable particles for
this taking into account their penetrating ability, which allow
them to interact with all the substance of the macroscopic
body --- not with the surface layer only. As is known~\cite{gt},
already in 30th Gamow and Teller offered to use the neutrinos for
the explanation of the gravitation, but their mechanism provided
the direct exchange of the pairs consisting of a neutrino and an
antineutrino and therefore the one does not correspond to the
modern conceptions of the theory of interactions.

Bashkin's works appearing in 80th on a propagation of the
spin waves in the polarized gases~\cite{ba} allowed to make the
supposition~\cite{k2}, that the analogous collective oscillations are
possible under certain conditions as well as in the neutrinos
medium. Since the collective oscillations can induce an
interaction between particles, Bashkin's works make us to pay
attention to the relict neutrinos~\cite{bv} (under which we shall imply
antineutrinos, too) filling our Universe. The effective
temperature~$T \approx 1.9~K \approx 1.64 \cdot 10^{-13}~GeV$
of the relict neutrinos is the fairly low one so that it is
fulfilled one of conditions ($ \lambda \gg r_w$, where~$\lambda$
is the de Broglie's wave-length of a neutrino and~$r_w$ is the
weak interaction radius of an one~\cite{ba}) of the propagation of the
spin wave in the polarized gases. As a result the quantum effects
become the determing ones in such medium and the interference of
the neutrinos fields (being the consequence of the known identity
of elementary particles) must induce the quantum beats, which will
be interpreted as zero oscillations of a vacuum. In consequence of
this the mathematical apparatus~\cite{gm} applied by the
description of the Casimir's effect~\cite{ca} can be used.

We shall be interesting in quantum beats arising by the
interference of the falling polarized flow of the relict neutrinos
 on the macroscopic body with the scattered one at this
body. Let's suppose for this the neutrinos have the zero rest
mass (the other version~\cite{bv} will not be considered), so that
the direction of their spin is connected hardly with the direction
of their 3-velocity. In consequence of this only those neutrinos
can be considered as ones forming the polarized flow,
which propagate along straight line connecting specifically
two particles of different macroscopic bodies. It explain the
anisotropy of the zero quantum oscillations, which is necessary
to  obtain  the right dependence~($1/R$) of the energy of the
two-particles interaction on the distance~$R$ between particles
in the Casimir's effect.

Let's consider two macroscopic bodies with masses~$m_1$
and~$m_2$ and with the fairly long distance~$R$ one from
another. We shall  regard,  that the bodies
contain~$2m_1 l$ and~$2m_2 l$ particles
correspondingly (where the normalizing factor~$l$ is
connected with cross-section~$\sigma $ of the neutrino upon
the particle), implying
thereby the statistics averaging of the properties of the
elementary particles constituting the bodies.  If the
particles of the macroscopic bodies had interacted with all
neutrinos incidenting on them then these particles might have been
considered as the opaque boandaries, which induce Casimir's
effect on the straight line. By this the energy of the
interaction of the particles would have been equal to~\cite{gm}
$$
\varepsilon_{AB} = \frac{1}{2}
  \sum_{n=1}^{\infty} \frac{\pi n}{R_{AB}} - \frac{1}{2}
  \int\limits_0^{\infty} \frac{\pi x}{R_{AB}} dx =
$$
\begin{equation}
 \label{1}
=\frac{i}{2} \int\limits_0^{\infty}
\frac{\pi (it)/R_{AB} - \pi (-it)/R_{AB}}{\exp (2\pi t) - 1}
 dt = - \frac{\pi}{24 R_{AB}}
\end{equation}
 ($A$ is a number of a particle of the first macroscopic
 body and~$B$ is a number of a particle of the second body).
 On account of the weakness of the  interaction of neutrinos
 with particles we are confined to a first approximation,  so that
 the energy~$E$ of the interaction of two macroscopic bodies is equal to
\begin{equation}
 \label{2}
 E\approx\sum_{A=1}^{2m_1 l}\sum_{B=1}^{2m_2 l}
 \varepsilon_{AB}.
\end{equation}
 Neglecting the dimensions of the bodies in comparison with interval
 $R$ between them (~$R_{AB}\approx R$), we shall  have finally
\begin{equation}
 \label{3}
 E\approx -2m_1 l~2m_2 l~\pi /(24R)
  = -G_{\nu} m_1 m_2 / R
\end{equation}
 where $ G_{\nu} = \pi l^2 /6 $.

\section{\hspace{-4mm}.\hspace{2mm} The estimate of the constant
 $G_{\nu}$}

 Consider the scattering of the neutrino upon the charge
 lepton, induced by the exchange of the neutral~$Z^{\circ}$~boson
 (taking account of the low energy of the relict neutrinos) only.
 The ammplitude of the process in the lower approximation can be
 written down as
$$
 M = 4G_F2^{-\frac12}(\nu_L^+\gamma_4\gamma^i\nu_L)[(-\frac12+\xi)
 e_L^+\gamma_4\gamma_ie_L + \xi e_R^+\gamma_4\gamma_ie_R] =
$$
$$
 = G_F2^{-\frac12}[\nu^+\gamma_4\gamma^i(1-\gamma_5)\nu]
 [(-\frac12+\xi)e^+\gamma_4\gamma_i(1-\gamma_5)e+
$$
\begin{equation}
 \label{1}
 +\xi e^+ \gamma_4\gamma_i(1+\gamma_5)e],
\end{equation}
 in consequence of this the square of the amplitude (spin-average) will
 take the form
$$
 <M^2>=64G_F^2[(-\frac12+\xi)^2(p'\cdot k')(p\cdot k)+
$$
\begin{equation}
 \label{2}
 +\xi^2(p'\cdot k)(p\cdot k')-(-\frac12+\xi)\xi m^2(k'\cdot k)].
\end{equation}
 where $\gamma^i$, $\gamma_i$ are the Dirac's matrices, $e$
 is a bispinor describing of a charge lepton, $p$ is its original of
 4-momentum and $p'$ is the finite 4-momentum of it ($m$ is the rest
 mass of a charge lepton); $\nu$ is bispinor, describing of the neutrino,
 $k$ is its original 4-momentum and $k'$ is the finite 4-momentun of it;
 $+$ is the symbol of the Hermitian conjugation;
 {$G_F \approx 1.166\cdot 10^{-5}~GeV^{-2}$} is Fermi's constant. Here
 and further
$$
 \xi =\sin^2\Theta_W,\quad\gamma_5=-i\gamma_1\gamma_2\gamma_3\gamma_4,
$$
\begin{equation}
 \label{3}
 \psi_L=\frac12(I-\gamma_5)\psi,\quad\psi_R=\frac12(I+\gamma_5)\psi.
\end{equation}
 ($I$ is the unit matrix and $\Theta_W$ is the weak angle).
 By analogy we can get the square of the scattering amplitude
 of the antineutrino upon the charge lepton as
$$
 <M^2>=64G_F^2[(-\frac12+\xi)^2(p'\cdot k)(p\cdot k')+
$$
\begin{equation}
 \label{4}
 \xi^2(p'\cdot k')(p\cdot k)-(-\frac12+\xi)\xi m^2(k\cdot k')].
\end{equation}

 As a result the cross-section of the scattering for the
 neutrino proves to be equaal to the cross-section of the
 scattering for the antineutrino in the low-energy
 approximation (the energy of the neutrino~$\omega \ll m$)
 and they are written down as
\begin{equation}
 \label{5}
 \sigma^Z=4G_F^2\omega^2(\frac14-\frac{\xi}2+\xi^2)/\pi .
\end{equation}
 Note that~$\sigma^Z$ proves to be minimal for~$\xi=1/4$.
 As the low energy neutrinos scarcely are able to change the
 spin direction of the particles of a macroscopic body,
 their scattering must be accompanied the collision
 radiation. In consequence of this the cross-section has
 the form
\begin{equation}
 \label{6}
 \sigma_{\nu} = k_{\psi}\sigma^Z,
\end{equation}
 where for the charge leptons the factor~$k_{\psi}=k_e$
 depends on the fine structure
 constant~$\alpha \approx 1/137$ only, while for the
 quarks the factor~$k_{\psi}=k_q$ must depends on the
 running coupling constant~$\alpha_s$ too, which define
 the collision radiation by gluons.

For the crude estimate of the constant~$G_{\nu}$ let us
consider the scattering the relict neutrino upon
the electron only, supposing that
\begin{equation}
 \label{7}
 \sigma_{\nu} = \pi l^2 ,
 k_e=\alpha(\pi^2-\frac{25}4)/(2\pi) .
\end{equation}
Besides substituting the middling
\begin{equation}
 \label{8}
<\omega >=\frac{\int\limits^{\infty}_0\omega^3 d\omega/
\left[\exp \left(\frac{\omega}T\right)+1\right]}
{\int\limits^{\infty}_0\omega^2 d
\omega/\left[\exp(\frac{\omega}T)+1\right]}
\end{equation}
instead of~$\omega$ we receive the following value
of the constant
\begin{equation}
 \label{9}
G_{\nu} =\sigma_{\nu}/6 \approx 10^{-38}~GeV^{-2}
\end{equation}
($<\omega > \approx 3.15 T \approx 5.166\cdot 10^{-13} GeV,
\xi \approx 0.23$) which is near to the known value of the
gravitational constant~$G_N$~\cite{k3}.

 So the gravitational phenomena can be explained by the presence of
 the collective oscillations in the neutrinos medium. In
 consequence it might be worthwhile to return to the potential
$$
 V(R) = \frac{A}{R} e^{-BR}
$$
 of which Seeliger~\cite{see} suggested to substitute the
 Newton's potential and to note the gravitational potential
 (it is possible in an any approximation) as
\begin{equation}
 \label{10}
 V(R) = \frac1R \sum_{i=1}^n A_i e^{-B_i R}
\end{equation}
 in the general case where the constants~$A_i$ and~$B_i$
 characterize the different media. By this we can be based on the
 theory of the strong gravity (see, for example the work~\cite{pea}).
 Moreover, having the neutrinos Universe and taking account of the
 Fermi-Dirac statistics we can recollect about the Saharov's
 hipothesis~\cite{sah}  using the idea of the metrical vacuum
 elasticity for the explanation of the gravitational interactions.
 But the main idea is it now for us what the normal matter (not
 neutrinos) acts as the Brownians by the help of which it can
 make the attempt to estimate the statistics characterization of
 the Universe neutrinos background. In the capacity of one from
 such indicator we offer to use the particles masses whiches
 connect with the scattering cross-section of the neutrinos.
 Note in tie with it, what we can ignore the photon collision
 radiation by the neutrinos scattering on the hadrons whiches
 are the quark resonator because of the existence of the additional
 degree of freedom in comparison with the electron. Exactly the
 resonance scattering causes to a gain in the hadrons masses by a
 factor of~$10^3$ in comparison with the electron mass. (The great
 spread of the hadrons masses depend on the form of the collective
 quark oscilations in the hadrons resonator.)

\section{\hspace{-4mm}.\hspace{2mm} The gauge transformations
                                    of the Lie local loop}

In order  to  get the more general equations than Einstein's
gravitational equations, one make use of the unified  covariant
gauge formalism~\cite{k5}. This formalism is differing from
the other ones, because gauge fields are being described by geometrical
objects of a tensor type but not by the connections, which
perform an other role and which may depend on gauge fields only
inderectly. Specifically, the application of connections as
counterterms~\cite{k5} is co-ordinating with the relativistic approach,
when the physical sence may be attached to a difference
of connections only~\cite{ch}. Besides we need geometrical  objects
describing  the ground state  of  a matter  (a vacuum),
which plays a role of the peculiar thermostat,  and connections
of various fiber bundles  suit for this splendidly. While using
connections, corresponding to the spaces of the rather complicated
geometrical structure, then one may consider that an excitation
state differ from a ground state a little. It allows to
consider Lagrangians,  depending on the lowest powers of covariant
derivatives only.  As a result  equations of gauge fields will
be generalization  of Maxwell's equations,  but one may obtain
in the particular case,  which  is considered  as the limiting
one, that they had a quasi-einstein form~\cite{k6} (when all
excitations are being freezed out).

Let's assume  that  Universe had a stage of a development
during which CPT-invariance of physical fields are absent. As
known~\cite{le} this situation must arise by a breakdown of Lorentz
invariance of the space-time~$M_4$ in which only the fields with
the  elemental  structure (scalar fields) can take place.
In consequence of physical processes unknown for us this stage
of  a  development was completed by a degeneracy of scalar fields
that led to their mixing and breaking down on classes of
an equivalence. As a result the laws of the quantum statistics
connected directly with the exchange-type interactions have started
playing the perceptible role.  Note that the exchange-type interactions
will not differ from other interactions formally in
our  approach and they are only a relict of primary ones after
a forming of spinor and vector fields. It is a degeneracy
that allows to use vector fiber bundles with the base~$M_4$.
Specifically, the fields~$\Psi(x)$ will be the cross sections of
the vector fiber bundle~$E_{4+N}$ (a point~$x\in M_4$). Using
of approximate symmetries by a description of interactions made
it possible to unite non-degenerate fields (in a general case)
in multiplets or in supermultiplets,  in  consequence  of
what a number~$N$ (in an abstract theory) is not concretized.

In addition to the physical fields must be described
by not  one  field~$\Psi (x)$,  but by the class of the
equivalence~$\{\Psi (x)\}$ in which the relation of the equivalence
is determined by the infinitesimal transformations having the form:
\begin{equation}
 \label{1}
\Psi\to\Psi +\delta\Psi =\Psi + \delta\omega^a~T_a\Psi ,
\end{equation}
where~$a,b,c,d,e = 1,2,...,r$; $\delta\omega^a(x)$ are infinitesimal
parameters; $T_a(x)$ are~$N\times N$~matrices depending on charges of
particles being quanta of fields~$\Psi (x)$. Since it is
  impossible to be fully
confident that there is a strict  border  between  internal
symmetries and external ones, then it is necessary to consider
both transformations of fields~$\Psi (x)$ and transformations
of points~$x$ in the form:
\begin{equation}
 \label{2}
x^i\to x^i+\delta x^i=x^i+\delta\omega^a~\xi_a^i(x),
\end{equation}
($x^i$ are coordinates of a point~$x\in M_4$;
$i,j,k,l,m,n,p,q = 1,2,3,4$).
In consequence of this it might be worthwhile to
resolve~$\delta\Psi$ into summands as follows
\begin{equation}
 \label{3}
\delta\Psi =\delta_0\Psi
+\delta\omega^a~\xi_a^i~\partial_i\Psi
\end{equation}
($\partial_i\Psi$ are the partial derivatives of
fields~$\Psi (x)$), selecting the
changes~$\delta_\circ \Psi$ of fields~$\Psi (x)$
in the point~$x$. Writing down~$\delta_0\Psi$ as
\begin{equation}
 \label{4}
\delta_0\Psi =\delta\omega^a~X_a(\Psi) =
\delta\omega^a~(T_a\Psi - \xi_a^i~\partial_i\Psi),
\end{equation}
we shall  regard~$X_a(\Psi)$ the generators of the Lie local
loop~$G_r$. We refuse the associativity property which is inherent
to the Lie local groups (see for example the work~\cite{kik}
of M.Kikkawa about a connection
of a geometry of a space with a structure of a Lie local loop).

Being on the relativistic positions it is necessary to  assume
that the change~$\delta\Psi$ can be only with the finite
rate and they are produced by an exchange of particles
or quasiparticles being quanta of
the special fields~$B(x)$,  the connection of which with the
fields~$\Psi (x)$ (in Lagrangian) must have the form:
\begin{equation}
 \label{5}
D_{\beta}\Psi =-B_{\beta}^aX_a(\Psi),
\end{equation}
where~$B_{\alpha}^a(x)$ are components of fields~$B(x)$ with a
consideration of their factorization on the Lie local loop~$G_r$.
Note, that~$B_{\alpha}^a(x)$ can be both Utiyama's gauge
fields~\cite{ut} and Kibble's gauge fields~\cite{kib}.
Following for Utiyama~\cite{ut} we shall not
   concretize
classes of an equivalence of fields~$B(x)$, in consequence
  of what  we shall
not concretize significances which are adopted
by indexes $\alpha$, $\beta$, $\gamma$, $\delta$, $\varepsilon$,
$\theta$. Resolving the matrices~$T_a$
into summands as
\begin{equation}
 \label{6}
T_a=L_a-\xi_a^i~\Gamma_i
\end{equation}
we write down~$X_a(\Psi)$ in the form:
\begin{equation}
 \label{7}
X_a(\Psi)=L_a\Psi-\xi_a^i~\nabla_i\Psi,
\end{equation}
where~$\nabla_i$ are symbols of covariant derivatives.

 Of course the connections~$\Gamma_i$ (as and any different ones)
 can considered as the gauge fields too, but only if we shall call
 the coordinate transformations as the gauge ones, which (in our
 opinion) must characterize the space of the observer with his
 instruments (containing the primary standards) and also the his
 method of the description (the physical model and the mathematical
 formalism).

  Let the type of geometrical objects to be conserved by the
 transformations of the Lie local loop~$G_r$ for what it is enough
 to demand that~$L_a(x)$ and~$\xi_a^i(x)$ are the linear homogeneous
 geometrical objects satisfying to the following relations:
\begin{equation}
 \label{8}
\xi_a^i~\nabla_i\xi_b^k - \xi_b^i~\nabla_i\xi_a^k
- 2~S_{ij}^k~\xi_a^i~\xi_b^j = - C_{ab}^c~\xi_c^k,
\end{equation}
\begin{equation}
 \label{9}
L_aL_b - L_bL_a - \xi_a^i~\nabla_iL_a + \xi_b^i~\nabla_iL_a
+ R_{ij}~\xi_a^i~\xi_b^j =  C_{ab}^c~L_c,
\end{equation}
where~$S_{ij}^k(x)$ are the components of the torsion of the
space-time~$M_4$
\begin{equation}
 \label{10}
 S_{ij}^k = (\Gamma_{ij}^k - \Gamma_{ji}^k)/2
\end{equation}
 and~$R_{ij}(x)$ are the components
 of the curvature of the connection~$\Gamma_i(x)$
\begin{equation}
 \label{11}
 R_{ij} = \partial_i \Gamma_j - \partial_j \Gamma_i +
 \Gamma_i \Gamma_j - \Gamma_j \Gamma_i .
\end{equation}
The components~$C_{ab}^c(x)$ of the
structural tensor of the Lie local loop~$G_r$ must satisfy
to the identities:
\begin{equation}
 \label{12}
C_{ab}^c + C_{ba}^c = 0,
\end{equation}
\begin{equation}
 \label{13}
C_{[ab}^d~C_{c]d}^e + \xi_{[a}^i~\nabla_{|i|} C_{bc]}^e
- R_{ij[a}{}^e~\xi_b^i~\xi_{c]}^j = 0,
\end{equation}
where~$R_{ija}{}^e(x)$ are the components of the curvature of the
connection~$\Gamma_{ia}{}^b(x)$
\begin{equation}
 \label{14}
 R_{ijb}{}^a = \partial_i \Gamma_{jb}{}^a -
 \partial_j \Gamma_{ib}{}^a + \Gamma_{ic}{}^a \Gamma_{jb}{}^c
 - \Gamma_{jc}{}^a \Gamma_{ib}{}^c .
\end{equation}

\section{\hspace{-4mm}.\hspace{2mm} The gauge fields}

Now one  may  proceed  to a construction of the covariant
gauge formalism following to R.Utiyama's classical work~\cite{ut}.
For  this it is necessary to find a law of a transformation of
the fields~$B(x)$. Let the fields~$D_{\alpha} \Psi$ change
analogously to the fields~$\Psi(x)$ in a point~$x\in M_4$, then is
\begin{equation}
 \label{1}
\delta_0 D_{\alpha}\Psi =\delta\omega^b~(L_bD_{\alpha}\Psi
- D_{\beta}\Psi~L_{b\alpha}{}^{\beta}
-\xi_b^i~\nabla_i D_{\alpha}\Psi),
\end{equation}
where the components~$L_{a\alpha}{}^{\beta}(x)$ of linear
homogeneous geometrical objects satisfy the following relations:
\begin{equation}
 \label{2}
L_{d\beta}{}^{\gamma}~L_{b\alpha}{}^{\beta} -
L_{b\beta}{}^{\gamma}~L_{d\alpha}{}^{\beta} -
\xi_d^i~\nabla_i L_{b\alpha}{}^{\gamma} +
\xi_b^i~\nabla_i L_{d\alpha}{}^{\gamma} +
R_{ij\alpha}{}^{\gamma}~\xi_d^i~\xi_b^j =
C_{db}^c~L_{c\alpha}{}^{\gamma}
\end{equation}
($R_{ij\alpha}{}^{\gamma}(x)$ are the components of the curvature
of the connection~$\Gamma_{i\alpha}{}^{\gamma}(x)$). As a
result~$\delta_0 B_{\alpha}^a$ are written down in the form:
\begin{equation}
 \label{3}
\delta_0 B_{\alpha}^d =
\delta\omega^b~(C_{cb}^d~B_{\alpha}^c -
B_{\beta}^d~L_{b\alpha}{}^{\beta} -
\xi_b^i~\nabla_i B_{\alpha}^d) +
\Phi_{\alpha}^i~(\nabla_i \delta\omega^d -
V_i~\delta\omega^d),
\end{equation}
where
\begin{equation}
 \label{4}
\Phi_{\beta}^i = B_{\beta}^a~\xi_a^i,
\end{equation}
\begin{equation}
 \label{5}
V_i = \Gamma_{ij}{}^j - \Gamma^o{}_{ij}{}^j.
\end{equation}
Here and further~$\Gamma_{ij}{}^k(x)$ are the components of the
internal connection of the space-time~$M_4$
and~$\Gamma^o{}_{ij}{}^k(x)$ are the components of the connection
of an equiaffine space of an affine connection being a locally
diffeomorphic one to the space of an affine connection~$M_4$.
As~$\delta_0 B_{\alpha}^c$ depend
on~$\nabla_i\delta\omega^a$ then~$B(x)$
are called the gauge fields.

Since the action
\begin{equation}
 \label{6}
\int\limits_{\Omega_4} {\cal L}\eta dx^1 dx^2 dx^3 dx^4
\end{equation}
($\Omega_4$ is a region of the space-time~$M_4$
and~$\eta (x)$ is the base density of the same) must be invariant
against infinitesimal transformations of the Lie local
loop~$G_r$, then the total Lagrangian~${\cal L}$ depending on
fields~$\Psi (x)$,~$B(x)$ and also their derivatives of
the first order is unable to be selected arbitrarily. The
following Lagrangian~${\cal L}(\Psi ;
D_{\alpha}\Psi; F_{\alpha\beta}^c)$ satisfy to this demand,
where the components~$F_{\alpha\beta}^c(x)$ of the intensities
of the gauge fields~$B(x)$ have the form:
$$
F_{\alpha\beta}^c =[\delta_b^c -
\xi_b^i~\Phi_i^{\gamma}~(B_{\gamma}^c -
\beta_{\gamma}^c)]~[\Phi_{\alpha}^j~\nabla_j B_{\beta}^b -
\Phi_{\beta}^j~\nabla_j B_{\alpha}^b - \\
$$
\begin{equation}
 \label{7}
B_{\alpha}^e~B_{\beta}^d~C_{ed}^b +
(B_{\alpha}^e~L_{e\beta}{}^{\delta} -
B_{\beta}^e~L_{e\alpha}{}^{\delta})~B_{\delta}^b].
\end{equation}
Note that the fields~$\Phi_i^{\alpha} (x)$ are defined
from the equations: $\Phi_{\alpha}^i~\Phi_j^{\alpha} =\delta_j^i$
($\delta_i^j$ and~$\delta_a^b$ are the Kronecker delta symbols).
The components~$\beta_{\alpha}^b (x)$ of linear homogeneous
geometrical objects are being interpreted as vacuum averages of
gauge fields~$B(x)$.

Rewrite the equations
\begin{equation}
 \label{8}
\Phi_{\alpha}^i~\left(\frac{\cal L}{\eta}~\frac{\partial
\eta}{\partial B_{\alpha}^b} + \frac{\partial {\cal L}}
{\partial B_{\alpha}^b} -
\nabla_j\left(\frac{\partial {\cal L}}
{\partial\nabla_j B_{\alpha}^b}\right)\right) = 0
\end{equation}
of gauge fields in the quasi-maxwell form:
\begin{equation}
 \label{9}
\nabla_j H_a^{ji} - H_a^{jk}~S_{jk}^i = I_a^i,
\end{equation}
where
\begin{equation}
 \label{10}
H_a^{ij} = -\Phi_{\beta}^i~\frac{\partial{\cal L}}
{\partial\nabla_j B_{\beta}^a} = \Phi_{\beta}^j~
\frac{\partial{\cal L}}{\partial\nabla_i B_{\beta}^a},
\end{equation}
\begin{equation}
 \label{11}
I_a^i = - {\cal L}\xi_a^i -
\frac{\partial {\cal L}}{\partial\nabla_i\Psi}~X_a(\Psi) -
\frac{\partial {\cal L}}{\partial\nabla_i B_{\beta}^b}~
Y_{a\beta}^b(B),
\end{equation}
\begin{equation}
 \label{12}
Y_{a\gamma}^b(B) = C_{ca}^b~B_{\gamma}^c - B_{\beta}^b~
L_{a\gamma}^{\beta} - \xi_a^i~\nabla_i B_{\gamma}^b.
\end{equation}
We pick out from the equations of gauge fields folding them
with~$B_{\alpha}^b~\Phi_l^{\alpha}$ those which can will be
called the equations of fields~$\Phi_{\alpha}^i (x)$ and which
must substitute for Einstein's gravitational equations in a
general case. When all excitation will be freezed
out ($B_{\alpha}^b\to\beta_{\alpha}^b$) these equations must
transfer to equations of a ground state of a matter (a vacuum),
it is desirable to write down in the geometrized form
(a quasi-einstein form) for the conservation of the Einstein's
ideology.

\section{\hspace{-4mm}.\hspace{2mm} The generalized Einstein's
                       equations}

Consider the specifical case, when Lagrangian has the following
form
$$
{\cal L}_t = F_{\alpha\beta}^a F_{\gamma\delta}^b
\eta^{\beta\delta}[\kappa\xi_a^i\xi_b^j
(\eta^{\alpha\gamma} \eta_{\varepsilon\theta}
h_i^{\varepsilon} h_j^{\theta} + 2 h_i^{\gamma} h_j^{\alpha}
- 4 h_i^{\alpha} h_j^{\gamma}) +
$$
\begin{equation}
 \label{1}
\kappa_1 \eta_{cd} \eta^{\alpha\gamma}
(\delta_a^c - \xi_a^i h_i^{\varepsilon}\beta_{\varepsilon}^c)
(\delta_b^d - \xi_b^j h_j^{\theta} \beta_{\theta}^d) ] /4
+ {\cal L}(\Psi, D_{\alpha}\Psi) ,
\end{equation}
where~$\kappa$ and~$\kappa_1$ are the constant ones;
$\alpha, \beta, \gamma, \delta, \varepsilon, \theta = 1,2,3,4$;
$\eta_{ab}(x)$, $\eta_{\alpha\beta}$ and $\eta^{\gamma\delta}$
are the components of the symmetric  undegenerate tensor fields,
by this~$\eta_{\alpha\beta} \eta^{\gamma\beta}
= \delta_{\alpha}^{\gamma}$.
Besides  the fields~$h_{\alpha}^i(x)$ are defined in the form
\begin{equation}
 \label{2}
h_{\alpha}^i = \beta_{\alpha}^a \xi_a^i
\end{equation}
and the fields~$h^{\alpha}_i$ are defined from the following
equations~$h^{\alpha}_i h_{\alpha}^j = \delta_i^j$.
Moreover let
\begin{equation}
 \label{3}
g^{ij} = \eta^{\alpha\beta} \Phi_{\alpha}^i \Phi_{\beta}^j ,
\end{equation}
\begin{equation}
 \label{4}
L_{a\alpha}{}^{\beta} = \xi_a^i L_{i\alpha}{}^{\beta} ,
\quad L_{k\gamma}{}^{\alpha} \eta^{\gamma\beta} +
L_{k\gamma}{}^{\beta} \eta^{\gamma\alpha} =
\eta^{\alpha\beta} l_k ,
\end{equation}
where
\begin{equation}
 \label{5}
l_k = (4 S_{ki}^i + g^{ij} \nabla_k g_{ij})/2 .
\end{equation}
By this the fields~$g_{ij}$ are defined from the
correlations~$g_{ij} g^{kj} = \delta_i^k$.
 As a result it can receive the equations
$$
 g^{jl} R_{ikl}{}^i - \frac12 \delta_k^j g^{ml} R_{iml}{}^i +
$$
$$
 \frac12 \nabla_i(2 g_{kp} g^{il} g^{jq} S_{ql}^p -
 g_{lk} g^{jq} \nabla_q g^{il} + g_{kq} g^{im} \nabla_m g^{qj} +
 S_{il}^j g^{im} g^{ln} (\nabla_m g_{kn} + S_{mn}^p g_{kp}) +
$$
$$
 g^{ij} [\frac12 (\nabla_k l_i - \nabla_i l_k) -
 l_l S_{ik}^l - \frac12 \nabla_m (g_{in} \nabla_k g^{mn})
 \nabla_m (g^{ml} S_{kl}^n g_{in}) + \nabla_m S_{ik}^m +
$$
$$
 \frac14 \nabla_k g^{lm} \nabla_i g_{lm} +
 S_{mk}^l g^{mn} \nabla_i g_{ln} - \frac14 l_k (4S_{li}^l +
 g_{lm} \nabla_i g^{lm}) - S_{pi}^n g_{mn} \nabla_k g^{mp} -
$$
$$
 2 S_{mk}^l (S_{li}^m + g_{ln} g^{mp} S_{pi}^n)] +
 \frac18 \delta_k^j [4 \nabla_i \nabla_l g^{il} -
 2 g_{mn} \nabla_i g^{lm} \nabla_l g^{in} -
 g^{il} \nabla_i g^{mn} \nabla_l g_{mn} +
$$
$$
 l^i (4 S_{li}^l + g_{lm} \nabla_i g^{lm}) +
 4 S_{il}^m g^{ln} (2 S_{mn}^i + g_{mp} g^{iq} S_{qn}^p) -
 8 S_{pq}^n g^{pm} g^{qi} \nabla_i g_{mn}] =
$$
\begin{equation}
 \label{6}
 \frac{1}{2\kappa} [D_a^{ij} E_{ik}^a -
 \frac14 \delta_k^j D_a^{il} E_{il}^a +
 P^j \Psi D_k \Psi - \delta_k^j {\cal L}(\Psi, D_i\Psi)] ,
\end{equation}
 where
\begin{equation}
 \label{7}
 D_i \Psi = \Phi_i^{\alpha} D_{\alpha}\Psi =
 \nabla_i\Psi - B_i^a L_a\Psi ,
\end{equation}
\begin{equation}
 \label{8}
 P^k\Psi = \frac{\partial {\cal L}}{\partial D_k\Psi} =
 \Phi_{\alpha}^k\frac{\partial {\cal L}}{\partial D_{\alpha}\Psi} ,
\end{equation}
\begin{equation}
 \label{9}
 B_i^a = \Phi_i^{\alpha} B_{\alpha}^a ,
\end{equation}
\begin{equation}
 \label{10}
 E_{ij}^a = (\delta_b^a - \xi_b^k B_k^a) (\nabla_i B_j^b -
 \nabla_j B_i^b + B_i^c B_j^d C_{cd}^b) ,
\end{equation}
\begin{equation}
 \label{11}
 D_a^{ij} = \kappa_1 g^{ik} g^{jl} g_{ab} E_{kl}^b ,
\end{equation}
which can be used for the description of both the medium with
the amply wide spectrum of properties and the unsterile vacuum
``spoiled" by the presence of fields.

\end{document}